\documentclass[aps,prl,superscriptaddress,floatfix,twocolumn,longbibliography]{revtex4-2}

\usepackage{amsmath,dcolumn,bm,amsthm,tabularx,subfigure}
\usepackage[colorlinks=true,urlcolor=blue,citecolor=blue,linkcolor=blue]{hyperref}
\usepackage{mathrsfs,times}
\usepackage{bbold,dsfont}
\usepackage{graphicx,graphics,color}
\usepackage{mathtools,nicefrac}
\usepackage{slashed}
\usepackage{txfonts}
\usepackage{amsfonts,amssymb}
\usepackage{multirow}
\allowdisplaybreaks

\newcommand{\beq}{\begin{equation}}
\newcommand{\eeq}{\end{equation}}
\newcommand{\bea}{\begin{eqnarray}}
\newcommand{\eea}{\end{eqnarray}}

\begin{document}
%------------------------------------------------------------------------------------------------------------------------------------------------------------------------------------------------------------
\title{Nonuniform quadrupolar orders in the spin-$3/2$ generalized Heisenberg chain}

\author{Jie~Chen}
\email[Corresponding author.\\]{chenjie666@sjtu.edu.cn}
\affiliation{Key Laboratory of Artificial Structures and Quantum Control (Ministry of Education), School of Physics and Astronomy, Shenyang National Laboratory for Materials Science, Shanghai Jiao Tong University, Shanghai 200240, China}

\author{Shijie~Hu}
\email[Corresponding author.\\]{shijiehu@csrc.ac.cn}
\affiliation{Beijing Computational Science Research Center, Beijing 100084, China}
\affiliation{Department of Physics, Beijing Normal University, Beijing, 100875, China}

\author{Lihui~Pan}
\affiliation{Key Laboratory of Artificial Structures and Quantum Control (Ministry of Education), School of Physics and Astronomy, Shenyang National Laboratory for Materials Science, Shanghai Jiao Tong University, Shanghai 200240, China}

\author{Xiaoqun~Wang}
\email[Corresponding author.\\]{xiaoqunwang@zju.edu.cn}
\affiliation{Key Laboratory of Artificial Structures and Quantum Control (Ministry of Education), School of Physics and Astronomy, Shenyang National Laboratory for Materials Science, Shanghai Jiao Tong University, Shanghai 200240, China}
\affiliation{School of Physics, Zhejiang University, Hangzhou 310058, Zhejiang, China}
\affiliation{Tsung-Dao Lee Institute, Shanghai Jiao Tong University, Shanghai 200240, China}
\affiliation{Collaborative Innovation Center of Advanced Microstructures, Nanjing University, Nanjing 210093, China}

\date{\today}
%------------------------------------------------------------------------------------------------------------------------------------------------------------------------------------------------------------
\begin{abstract}

The generation of nonuniform quadrupole states plays a crucial role in understanding various fascinating phenomena observed in the advancement of several research areas, e.g., multiferroic compounds, nonmagnetic superconductors, etc.
In this work, we investigate the ground-state phase diagram of a generalized spin-$3/2$ bilinear-biquadratic-bicubic Heisenberg chain in the representation of multipolar operators.
By numerical simulations with the large-scale density-matrix renormalization group (DMRG) method, we successfully identify a tetramerization phase and a stripe-Q phase.
These phases are characterized by the emergence of nonuniform quadrupole orders resulting from the spontaneous breaking of translation symmetry.
In particular, tetramerization phase refers to the quadrupole operators take a four-cycle, while the stripe-Q phase represents a striped pattern in quadrupole operators.
Additionally, we demonstrate the presence of a Wess-Zumino-Witten (WZW) model with level $k = 1$ at the transition point between the dimerized (DM) phase and the Luttinger liquid (LL) phase, based on strong numerical findings.

\end{abstract}
%------------------------------------------------------------------------------------------------------------------------------------------------------------------------------------------------------------
\maketitle
%------------------------------------------------------------------------------------------------------------------------------------------------------------------------------------------------------------
{\it \textcolor{blue}{Introduction}}---In recent years, there has been considerable attention focused on the intriguing spin-$S=3/2$ physics associated with nematicity, leading to exciting developments in various research areas.
One fascinating example is the compound $\text{Ba}_2\text{CoGe}_2\text{O}_7$, where the coupling between the Co$^{2+}$ ions multiple spin operators and the electric polarization on the ligands gives rise to exotic magnetic anisotropy and spin dynamics~\cite{Soda_2014}.
By partially substituting nonmagnetic $\text{Zn}$ for $\text{Co}$, the magnetic order dome transforms into a double dome structure due to the emergence of magnetic Bose-Einstein condensates occurring in three excitations~\cite{Watanabe_2023}.
Another noteworthy case is the half-Heusler semimetal $\text{YPtBi}$, where the pairing between spin-$3/2$ fermion states leads to a dominant septet pairing state, which is qualitatively different from the theory of pairing between spin-$1/2$ states~\cite{Brydon_2016, Kim_2018}.
In the realm of nonmagnetic superconductors, the presence of quadrupolar charge order in the lattices of $\text{FeSe}_{1-x}\text{S}_x$ compounds plays a crucial role in stabilizing a large superconducting gap and showcasing novel dynamics~\cite{Zhang_2021}.%
Moreover, epitaxially strained $\text{Cr}$-based monolayers, such as $3$d transition metal compounds like $\text{CrSiTe}_3$, propose a possible Kitaev spin liquid state, where $\text{Cr}$ possesses a spin value of $S=3/2$~\cite{Xu_2020}.

The pure spin-$3/2$ Heisenberg chain belongs to the same universality class as the spin-$1/2$ Heisenberg chain~\cite{Hallberg_1996}, confirming the Haldane conjecture~\cite{haldane19883}.
When considering degenerate orbitals at quarter-filling, nematic orders in the Mott-insulator can be effectively described by the symmetric Kugel-Khomskii (KK) model~\cite{Kugel_1982}.
This model combines two orbitals into an SU($4$) spin, which is relevant in transition-metal oxide/dichalcogenide~\cite{daghofer2008absence, wu2018hubbard}, organic molecular compound~\cite{janani2014haldane}, unconventional superconductors~\cite{dagotto2011properties, takimoto2004strong}, and ultracold atomic gases in the optical lattices~\cite{dutta2015non}.
Beyond the degenerate limit, models with SU($2$) symmetry for spin-$3/2$ on the bonds can be represented by a prototype known as the bilinear-biquadratic-bicubic (BBB) Heisenberg model~\cite{Yamashita98}.
The model Hamiltonian reads
\begin{eqnarray}
\hat{\mathcal{H}}=\sum_{\left< l,l' \right>}{\left[ J\left( \mathbf{\hat{S}}_l\cdot \mathbf{\hat{S}}_{l'} \right) +K\left( \mathbf{\hat{S}}_l\cdot \mathbf{\hat{S}}_{l'} \right) ^2+L\left( \mathbf{\hat{S}}_l\cdot \mathbf{\hat{S}}_{l'} \right) ^3 \right]}\,\,
\label{Ham}
\end{eqnarray}
where $\langle \rangle$ sums over all the possible links, parameters $J = \cos \theta \cos \varphi$, $K = \cos \theta \sin \varphi$ and $L = \cos \theta \cos \varphi$ are controlled by two angles $\theta$ and $\varphi$.
The presence of uniform spin (anti-)nematic states in a spin-$3/2$ isotropic non-Heisenberg magnet in two or three dimensions has been discovered~\cite{Fridman_2011}.
However, the following questions in a chain remain unclear:
(1) Does a specific commensurate or even incommensurate quadrupolar order emerge due to the spontaneous breaking of translation symmetry? The mechanism could provide valuable insight into the quadrupolar stripe order and dynamics of nonmagnetic superconductors in two dimensions.
(2) Is the regular spin-nematic order affected by the cubic terms in \eqref{Ham}? It is possible that an entirely new criticality paradigm is induced.

Combining the analytical redefinition of high-order spin operators with large-scale density-matrix renormalization group (DMRG) simulations, we find a tetramer phase and a stripe phase of the quadrupole operator. We carefully studied the dimer phase to the Luttinger liquid phase in the phase diagram. We illustrated that it could fit well with the previous conclusions of the CFT, which directly indicates that the effective model here is the WZW model with level $k=1$. 

%------------------------------------------------------------------------------------------------------------------------------------------------------------------------------------------------------------
{\it {\color{blue}Multipole operators and SU($4$) ULS points.}}---In the generalized Heisenberg model that includes all possible higher-order terms, there exists a series of high symmetry points known as Uimin-Lai-Sutherland(ULS) points\cite{Uimin70,Lai74,Sutherland75}.
The one-dimensional generalized Heisenberg chain with $S=1$ containing bilinear and biquadratic terms, $H_{S=1}=\sum_{\left< i,j \right>}{\left[ \cos \theta \mathbf{S}_i\cdot \mathbf{S}_j+\sin \theta \left( \mathbf{S}_i\cdot \mathbf{S}_j \right) ^2 \right]}$, has been well studied in much literature, and its phase diagram has been determined\cite{Fath91,Fath95,Hu14,Itoi97,Uimin70,Lai74,Sutherland75,AKLT87,Takhtajan82,Babujian82}, there exists an SU($3$) antiferromagnetic ULS point at $\theta = \pi/4$, and an SU($3$) ferromagnetic ULS point at $\theta = 5\pi/4$.

The so-called spherical tensor operators\cite{SuppMat} can be defined and are basis vectors for the irreducible representation of the rotation group; they can be combined to form the so-called multipole operators. We define the $l$-th-order spherical tensor operator as $T^l$, which contains three components $\left\{ T_{n}^{l},T_{0}^{l},T_{-n}^{l} \right\}$, where $n\in \left\{ 1,2...l \right\}$. The multipole operator can be obtained using the spherical tensor operator, which is represented as follows
\begin{eqnarray}
M_{0}^{l}&=&\sqrt{2}T_{0}^{l}\\
M_{n}^{l}&=&\left[ T_{-n}^{l}+\left( - \right) ^nT_{n}^{l} \right]\\
M_{-n}^{l}&=&i\left[ T_{-n}^{l}-\left( - \right) ^nT_{n}^{l} \right]
\end{eqnarray}
here the superscript $l$ corresponds to the angular momentum quantum number, and the subscript is the magnetic quantum number $m=\left\{ -n,0,n \right\} \in \left\{ -l...l \right\}$. Generally, one uses the notation in quantum mechanics to mark these multipole operators, for example, the first-order operator, $M^1=\left\{ M_{0}^{1},M_{1}^{1},M_{-1}^{1} \right\}$ is written as $\mathbf{S}=\left\{ S^z,S^x,S^y \right\}$, also known as the spin operators; the second-order operators $M^2=\left\{ M_{0}^{2},M_{1}^{2},M_{-1}^{2},M_{2}^{2},M_{-2}^{2} \right\}$ is written as $\mathbf{Q}=\left\{ Q^{3z^2-r^2},Q^{zx},Q^{yz},Q^{x^2-y^2},Q^{xy} \right\}$, this is the common quadrupole operators. Here we define the octupole operators as $\mathbf{G}=\left\{G^{z\left(5 z^2-3 r^2\right)}, G^{x\left(5 z^2-r^2\right)}\right.$ $\left.G^{y\left(5 z^2-r^2\right)}, G^{z\left(x^2-y^2\right)}, G^{x y z}, G^{x\left(x^2-3 y^2\right)}, G^{y\left(3 x^2-y^2\right)}\right\}$, they correspond to the third-order operators $M^3=\left\{M_0^3, M_1^3, M_{-1}^3, M_2^3, M_{-2}^3, M_3^3, M_{-3}^3\right\}$.

It can be shown\cite{SuppMat} that there are some equivalence relations between multipole operators and spin operators as follows
\begin{equation}
\mathbf{Q}_i \cdot \mathbf{Q}_j=\left(\mathbf{S}_i \cdot \mathbf{S}_j\right)+2\left(\mathbf{S}_i \cdot \mathbf{S}_j\right)^2-\frac{2}{3}S^2S^2
\end{equation}
\begin{eqnarray}
\mathbf{G}_i\cdot \mathbf{G}_j&=&-\frac{2}{5}\left( 3S^2S^2-2S^2-3 \right) \left( \mathbf{S}_i \cdot \mathbf{S}_j \right)+4\left( \mathbf{S}_i \cdot \mathbf{S}_j \right) ^2\nonumber\\
&&+2\left( \mathbf{S}_i \cdot \mathbf{S}_j \right) ^3-S^2S^2
\end{eqnarray}
here $S^2=S\left( S+1 \right)$. Using these equations, the Hamiltonian (\ref{Ham}) can be re-expressed in terms of multipole operators as follows
\begin{equation}
H_{S=3/2}=g_1(\mathbf{S}_i \cdot \mathbf{S}_j)+g_2(\mathbf{Q}_i \cdot \mathbf{Q}_j)+g_3(\mathbf{G}_i\cdot \mathbf{G}_j)+\text { const. }
\label{Ham1}
\end{equation}
over here,$g_1=\left[(\cos \theta \cos \varphi)-\frac{1}{2}(\cos \theta \sin \varphi)+\frac{1}{5}(\sin \theta) \times\right.\\$$\left.\left(3 S^2 S^2-2 S^2+2\right)\right]$, $g_2=\left[\frac{1}{2}(\cos \theta \sin \varphi)-(\sin \theta)\right]$, $g_3=\left[\frac{1}{2}(\sin \theta)\right]$, const. $=\left[\frac{1}{3}(\cos \theta \sin \varphi)-\frac{1}{6}(\sin \theta)\right] S^2 S^2$.

\begin{figure}[t]	
	\begin{center}
		
		\includegraphics[scale=0.52]{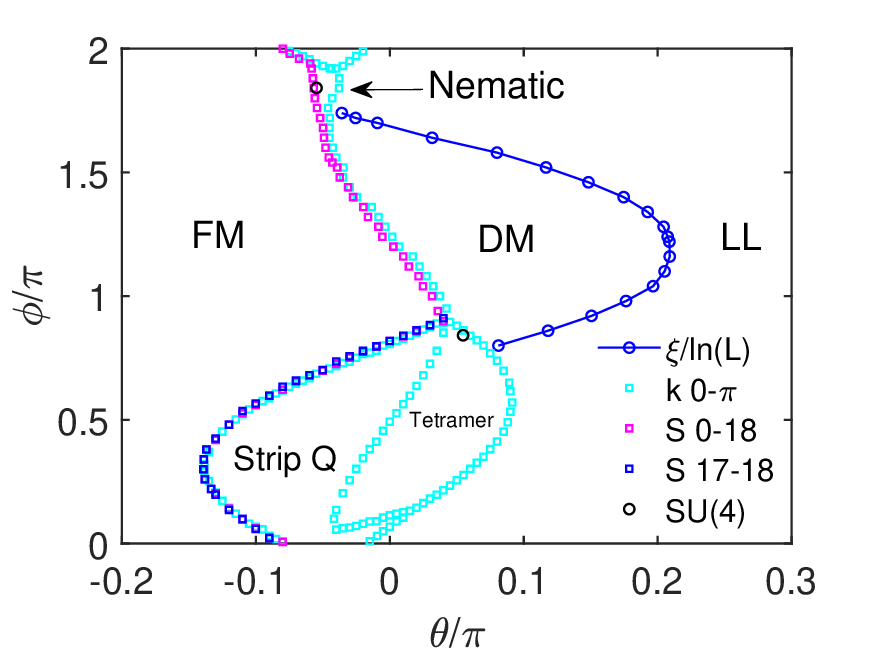}
		
		\caption{The phase diagram of the model, where FM denotes the ferromagnetic phase, DM denotes the dimer phase, LL denotes the Luttinger liquid phase, and stripe-Q denotes the stripe phase of the quadrupole operator. The determination of the phase transition boundaries is detailed in the main text.}
		\label{pic1}
		
	\end{center}
\end{figure}

We prove the following equation\cite{SuppMat}
\begin{equation}
\left(\boldsymbol{\lambda}_i \cdot \boldsymbol{\lambda}_j\right)=\frac{2}{5}\left(\mathbf{S}_i \cdot \mathbf{S}_j\right)+\frac{1}{6}\left(\mathbf{Q}_i \cdot \mathbf{Q}_j\right)+\frac{2}{9}\left(\mathbf{G}_i \cdot \mathbf{G}_j\right)
\end{equation}
here $\boldsymbol{\lambda}$ is the 15 SU($4$) generators, and combined with the above equation (\ref{Ham1}), the Hamiltonian can be rewritten as $H=g\left(\boldsymbol{\lambda}_i \cdot \boldsymbol{\lambda}_j\right) +const.$, where $g=\frac{5}{2} g_1=6 g_2=\frac{9}{2} g_3$, so the system has SU($4$) symmetry at this point, this corresponds to the KK model mentioned above. It can be found that the parameters at this point are $\varphi=\arctan \left(-\frac{44}{81}\right)$ and $\theta=\arctan \left(-\frac{16}{81} \cos \left(\arctan \left(-\frac{44}{81}\right)\right)\right)$. At this time there is $g=-36 / \sqrt{8753}$ and $H_{S=3 / 2}=\frac{1}{\sqrt{8753}} \sum_{\langle i, j\rangle}\left[81\left(\mathbf{S}_i \cdot \mathbf{S}_j\right)-44\left(\mathbf{S}_i \cdot \mathbf{S}_j\right)^2-16\left(\mathbf{S}_i \cdot \mathbf{S}_j\right)^3\right]+$ const. , the expressions in square brackets are the commonly known SU($4$) high symmetry point\cite{Yamashita98,Itoi97}. It is the SU($4$) ULS point at the top of the Fig.~\ref{pic1}, which is the SU($4$) ferromagnetic point because the $g$ of this point is negative; the other SU($4$) ULS point, whose $g$ is positive, corresponds to the SU($4$) antiferromagnetic point, which is known from the triangular relationship, corresponding to the one at the below of the Fig.~\ref{pic1}.
%------------------------------------------------------------------------------------------------------------------------------------------------------------------------------------------------------------

%---------------------------------------------------------------------------------------
%---------------------------------------------------------------------------------------

{\it {\color{blue}Phase diagram.}}---First, we calculated the ground state energy using the exact diagonalization(ED)\cite{sandvik2010computational} and the Density Matrix Renormalization Group(DMRG)\cite{schollwock2011density} with SU($2$) symmetry and determined the main phase boundaries, the main results are shown in Fig.~\ref{pic1}. We chose the chain length to be $L=12$ when its maximum total spin is $18$. As shown later, the leftmost region corresponds to the ferromagnetic phase, whose total spin is at its maximum value. So as shown in the Fig.~\ref{pic1}, we calculated the ground state energy for the two cases, $\left[E_0(S=17), E_0(S=18)\right]$ and $\left[E_0(S=0), E_0(S=18)\right]$,  and obtained their intersection points. Thus, the phase boundary of the ferromagnetic phase breakage can be obtained. It can be seen that the boundaries fixed in these two ways overlap in the lower half, which means that the phases at the top and bottom of the figure are not the same. We again calculated the case $\left[E_0(k=0), E_0(k=\pi)\right]$ with ED and also obtained the boundary of their intersection, which can be seen to fix a new phase in the lower part of the figure, and coincides with the SU($2$)-DMRG result at the ferromagnetic phase boundary, except for a small area at the top of the figure.

The phase transition boundary from the dimer phase to the Luttinger liquid phase in Fig.~\ref{pic1} is not obtained through the intersection of the ground state energies due to the strong finite size effect, and how the phase transition boundary in Fig.~\ref{pic1} is obtained will be discussed carefully later. To understand the nature of these phases, we have calculated the observed quantities in different phase regions using DMRG with $\sum_i S_i^z=0$ in the following three cases of $\phi / \pi=0.5$, $\phi / \pi=1.24$ and $\phi / \pi=1.8$, respectively.

%---------------------------------------------------------------------------------------
\begin{figure}[t]
	
	\centering
	
	\subfigbottomskip=0.1pt
	
	\subfigure{
		\begin{minipage}{0.5\linewidth}
			\centering
			\includegraphics[scale=0.31]{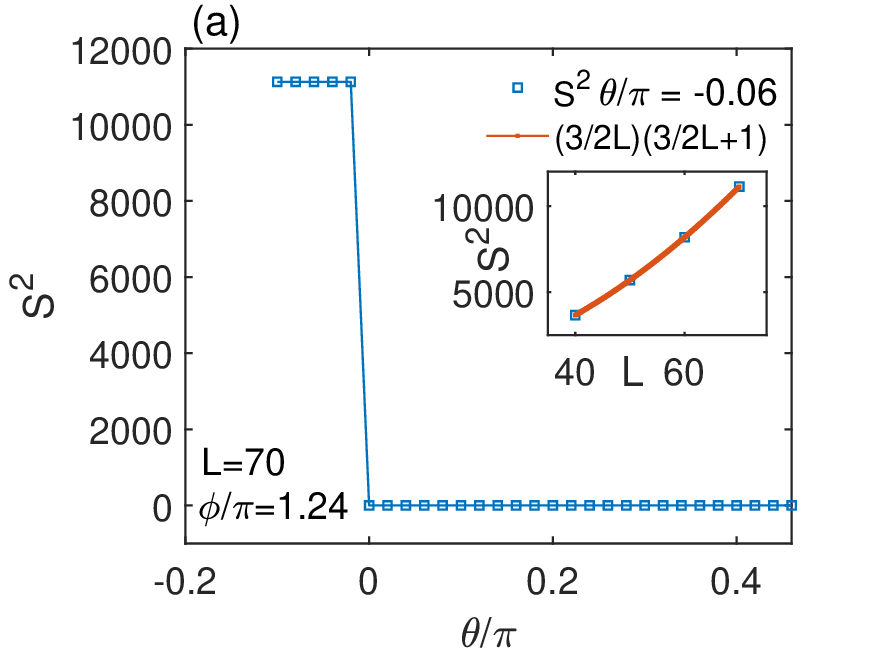}
			%\caption{fig1}
		\end{minipage}
	}%
	\subfigure{
		\begin{minipage}{0.5\linewidth}
			\centering
			\includegraphics[scale=0.31]{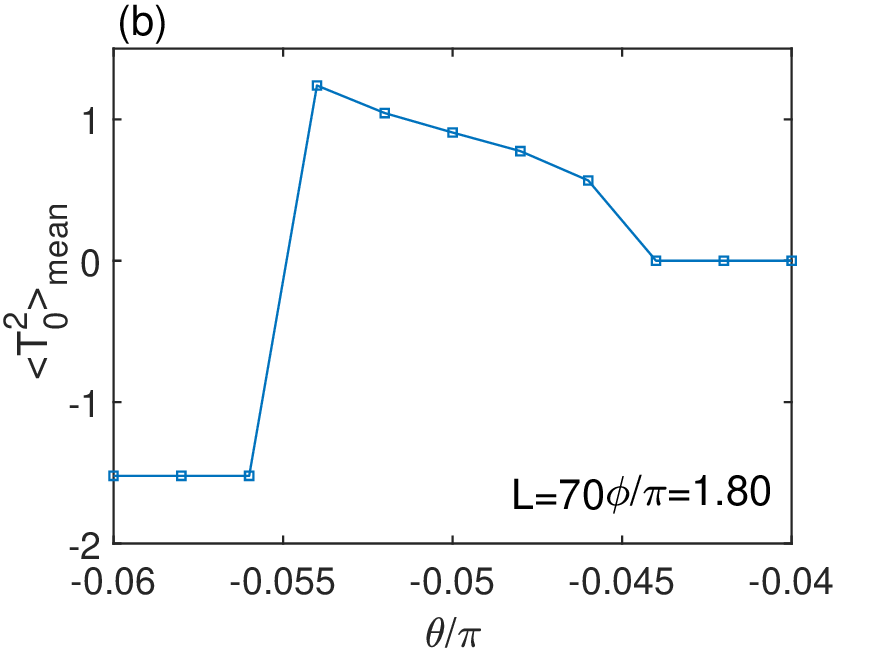}
			%\caption{fig2}
		\end{minipage}
	}%
	\quad
	\subfigure{
		\begin{minipage}{0.5\linewidth}
			\centering
			\includegraphics[scale=0.31]{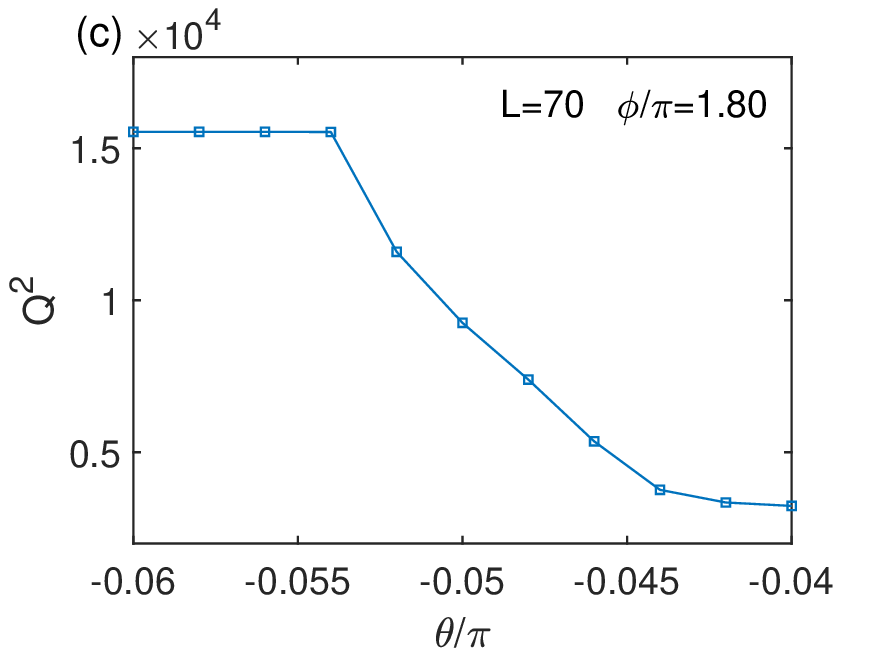}
			%\caption{fig3}
		\end{minipage}
	}%
	\subfigure{
		\begin{minipage}{0.5\linewidth}
			\centering
			\includegraphics[scale=0.31]{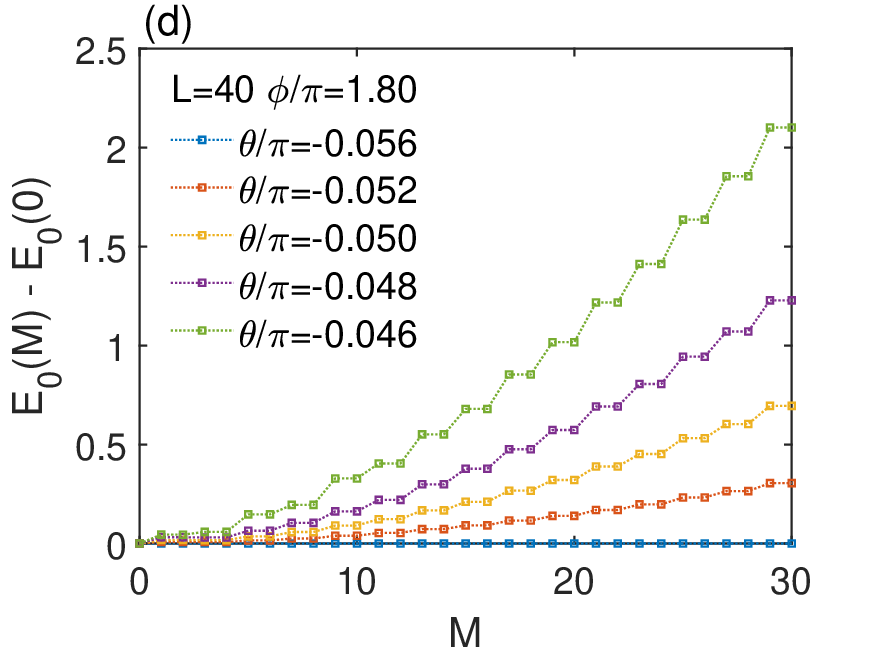}
			%\caption{fig4}
		\end{minipage}
	}%
	
	\caption{(a) Total $S^2$ values for $\phi / \pi=1.24$, chain length $L=70$, inset showing $S^2$ at its possible maximum value when $\theta / \pi=-0.06$ at different sizes. (b)$\phi / \pi=1.80$, the average value on each lattice site of $T^2_0$ at chain length $L=70$. (c) $\phi / \pi=1.80$, the total $Q^2$ value for chain length $L=70$. (d) The energy change of the ground state for a fixed different total magnetization $M=\sum_i S_i^z$ at chain length $L=40$.}
	\label{pic2}
	
\end{figure}
%---------------------------------------------------------------------------------------

As shown in the Fig.~\ref{pic2}(a), since the spin of the system is $\frac{3}{2}$, the maximum spin is  $\left(\frac{3}{2} L\right)\left(\frac{3}{2} L+1\right)$ for a chain of length $L$; from the inset, it can be seen that in the left region, the total spin is maximum; this indicates that the system is in the ferromagnetic phase.

We test the small region at $\phi / \pi=1.8$ and find a significant jump in the value of $T^2_0$ in the small region, as shown in Fig.~\ref{pic2}(b). We also calculate $S_z$ and $T^3_0$ and find them essentially $0$. This region is the nematic phase, for which the system's excitation is not a single magon but the bound state of two magnons. As shown in Fig.~\ref{pic2}(d), we test the variation of the ground state energy at $L=40$, with different $M=\sum_i S_i^z$. Indeed the system is in the small region where the ground state energy becomes larger with the interval $\triangle M=2$ stepwise, which corresponds to the excited state being two magons bound states. From the above equations it can be obtained that $Q \cdot Q=2\left[T_{i 2}^2 T_{j-2}^2-T_{i 1}^2 T_{j-1}^2+T_{i 0}^2 T_{j 0}^2-T_{i-1}^2 T_{j 1}^2+T_{i-2}^2 T_{j 2}^2\right]$, the calculation result of $Q^2$ is shown in Fig.~\ref{pic2}(c). In the FM region, its value is constant, and after the phase change point, its value gradually decreases in the small region.

\begin{figure}[b]
	
	\centering
	
	\subfigbottomskip=0.1pt
	
	\subfigure{
		\begin{minipage}{0.5\linewidth}
			\centering
			\includegraphics[scale=0.31]{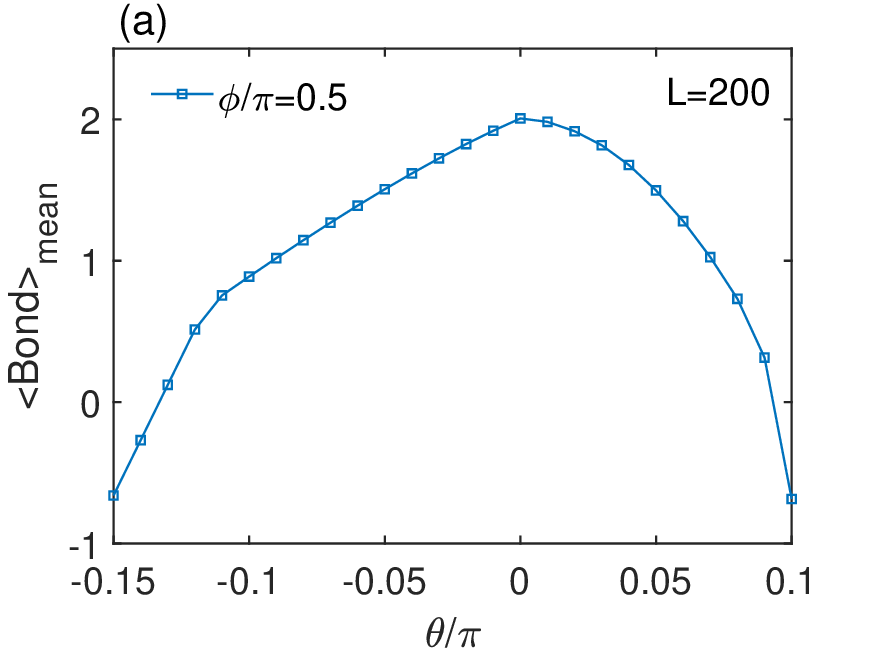}
			%\caption{fig1}
		\end{minipage}
	}%
	\subfigure{
		\begin{minipage}{0.5\linewidth}
			\centering
			\includegraphics[scale=0.31]{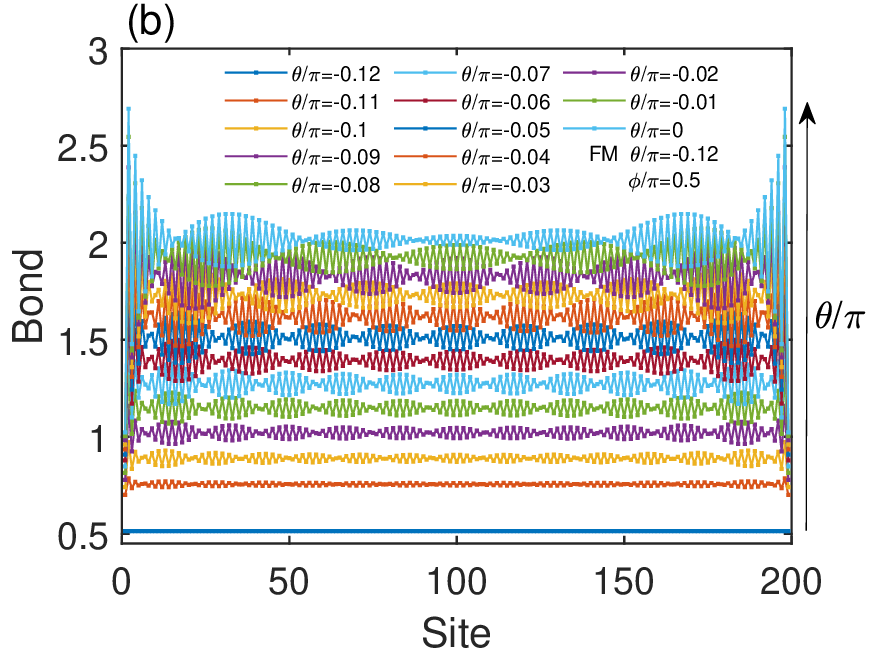}
			%\caption{fig2}
		\end{minipage}
	}%
	\quad
	\subfigure{
		\begin{minipage}{0.5\linewidth}
			\centering
			\includegraphics[scale=0.30]{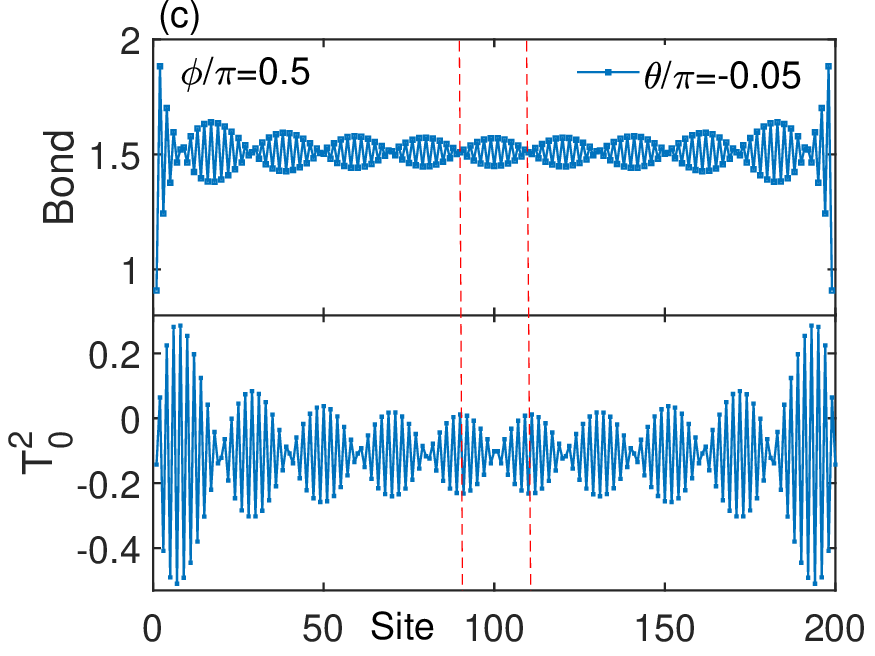}
			%\caption{fig3}
		\end{minipage}
	}%
	\subfigure{
		\begin{minipage}{0.5\linewidth}
			\centering
			\includegraphics[scale=0.30]{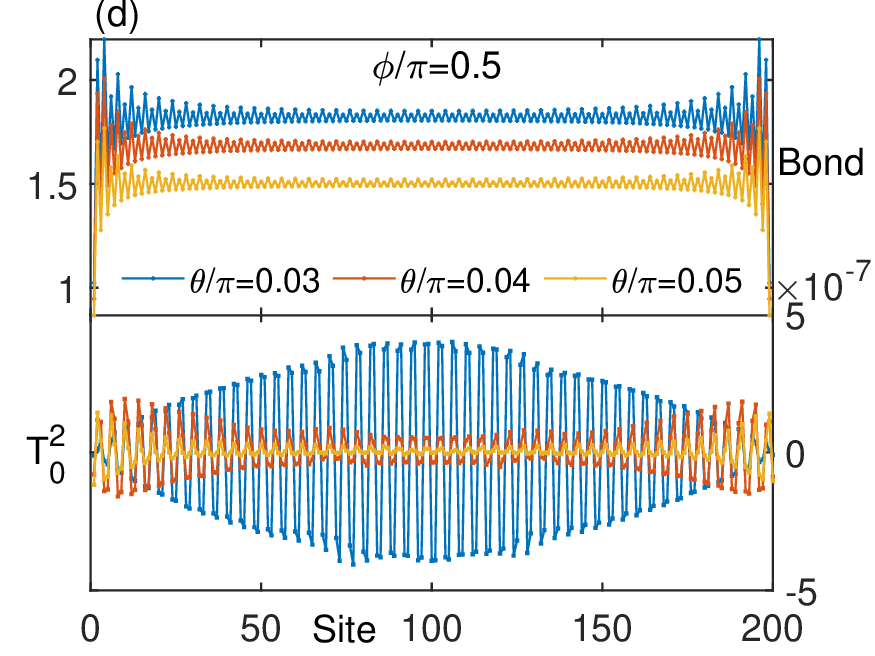}
			%\caption{fig4}
		\end{minipage}
	}%
	
	\caption{(a) The average energy on each bond for $\phi / \pi=0.5$ and $L=200$. (b) Variation of the fluctuation of energy on the bonds for $\phi / \pi=0.5$, $L=200$, as the parameter $\theta$ changes. (c) Comparison of the energy on the bond and the value of $\mathrm{T}_0^2$ on each lattice site for $\phi / \pi=0.5$, $\theta / \pi=-0.05$, and $L=200$. (d) Comparison of the energy on bond and the value of $\mathrm{T}_0^2$ on each site for $\phi / \pi=0.5$, $L=200$, $\theta / \pi=0.03, 0.04,0.05$.}
	\label{pic3}
	
\end{figure}

{\it {\color{blue}Stripe-Q and Tetramer phases.}}---In the following, we focus on the results when $\phi / \pi=0.5$ to illustrate the stripe phase of the quadrupole operator and tetramer phase we found. First, as shown in Fig. ~\ref{pic3}(a), we calculate the average energy on the bond, and we can see that at several phase transition points, there is a clear and sharp change in the bond energy, which will be consistent with the phase boundary determined by the ED in Fig.~\ref{pic1}.

In Fig.~\ref{pic3}(b), we carefully inspect the energy on the bond and find that after the ferromagnetic phase, the energy on the bond shows incommensurate stripes and its period becomes larger as $\theta$ becomes larger, i.e., the wave packets become larger, and their number becomes smaller. We calculate $S_z$, $T^2_0$ and $T^3_0$ on each lattice site and find that only $T^2_0$ also forms similar stripes, as shown in Fig.~\ref{pic3}(c). It can be seen that the peak of $T^2_0$ is just the trough of the bond, which predicts that the fluctuation of $T^2_0$ causes the stripes on the energy. This phenomena can be analyzed as follows, as shown in Fig.~\ref{pic3}(c); due to the interaction, the chain length of $L=200$ can be renormalized as the chain length that is reformed to $L=10$, then each site is put a $T^2_0$, they have a certain localization area. They can move freely in this area, but the $T^2_0$ between different lattice points have repulsive interactions; when they are close, the energy will increase. This explains the alignment between the wave peaks of the $T^2_0$ and the troughs of bond energy in Fig.~\ref{pic3}(c). This means we have a stripe phase of the quadrupole operator, which we call ‘stripe-Q’ phase.

%---------------------------------------------------------------------------------------
%---------------------------------------------------------------------------------------

If we continue to increase $\theta$, the system will enter a new phase when it passes the stripe-Q region, and the bond energy will take on a four-cycle nature, as shown in Fig.~\ref{pic3}(d), which we call it tetramer phase. The $T^2_0$ is small at this point, but the same four-cycle behavior can still be seen. When $\theta=0$, formally, at this point, the Hamiltonian quantities agree with the generalized Heisenberg model of $S=1$, it can be found that the phase diagrams of both are roughly consistent. Except that the Haldane phase in $S=1$ now becomes the Luttinger liquid phase, and there is an additional Stripe-Q phase region in $S=3/2$. The trimer in $S=1$ is now a tetramer. In $S=1$, the literature\cite{Itoi97} illustrates the trimer phase as critical using the non-abelian bosonization near the ULS point of the SU($3$) symmetry, whether our tetramer phase here is a critical phase value for further study.

{\it {\color{blue}Dimer to Luttinger liquid phase transition.}}---Affleck's work\cite{affleck1989critical} gives a general formula for the dimer energy gap $\Delta \propto g_i^{1 /\left(2-x_i\right)} /\left|\ln g_i\right|^{2 b_i / b\left(2-x_i\right)}$, here $g_i$ generally satisfies the renormalization equation $\mathrm{d} g_i / \mathrm{d} l=\left(2-x_i\right) g_i-2 \pi b_i g_i g$ \cite{cardy1996scaling,affleck1989critical,cardy1986logarithmic}, here $e^l=L$ and $L$ is the chain length, $g$ and $g_i$ are the coupling coefficients of the operator and the primary fields, respectively, and $b$ and $b_i$ are the coefficients of their three-point correlation functions. Here $2$ denotes the spacetime dimension, and $x_i$ is the conformal dimension; when $x_i>2$, the field is irrelevant, $x_i<2$ is relevant, and $x_i=2$ is marginal. For a general spin half-integer system, for a model whose effective theory is $H=\sum_i\left[1+(-)^i \delta\right] \mathbf{S}_i \mathbf{S}_{i+1}$, here $\delta$ is $g_i$, $2 b_i / b=\frac{3}{4}$, the conformal dimension $x=2s_l\left( s_l+1 \right) /\left( 2+k \right)$, here, $k$ is the level of the Kac-Moody algebra (WZW model), and the left-hand spin $s_l$ and the right-hand spin $s_r$ of the primary field operator takes values in $[0,1 / 2, \ldots, k / 2]$. When taking the WZW model with $k=1$, it corresponds to a free boson theory, which is a stable critical point\cite{affleck1989critical}. At this point, $x=3 / 2(2+k)=1 / 2$, so the energy gap is $\Delta \propto \delta^{2 / 3} /|\ln \delta|^{1 / 2}$. In general, the correlation length $\xi / l \sim \frac{1}{\Delta}$, so we get
\begin{equation}
\frac{\xi}{\ln L} \sim \frac{|\ln \delta|^{1 / 2}}{\delta^{2 / 3}}
\label{gap}
\end{equation}
recalling Eq. \ref{Ham1}, we have rewritten the Hamiltonian as a summation of three bilinear terms. At this point, there are similar relevant operators and primary fields in all three, which lead to energy gaps in all three as in Eq. \ref{gap}.

\begin{figure}[t]
	
	\centering
	
	\subfigbottomskip=0.1pt
	
	\subfigure{
		\begin{minipage}{0.5\linewidth}
			\centering
			\includegraphics[scale=0.31]{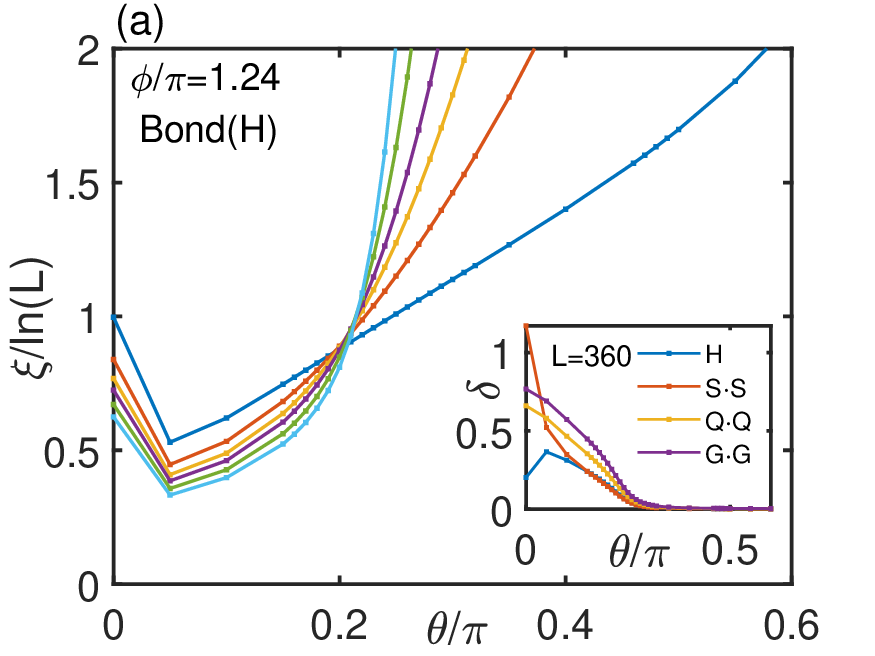}
			%\caption{fig1}
		\end{minipage}
	}%
	\subfigure{
		\begin{minipage}{0.5\linewidth}
			\centering
			\includegraphics[scale=0.31]{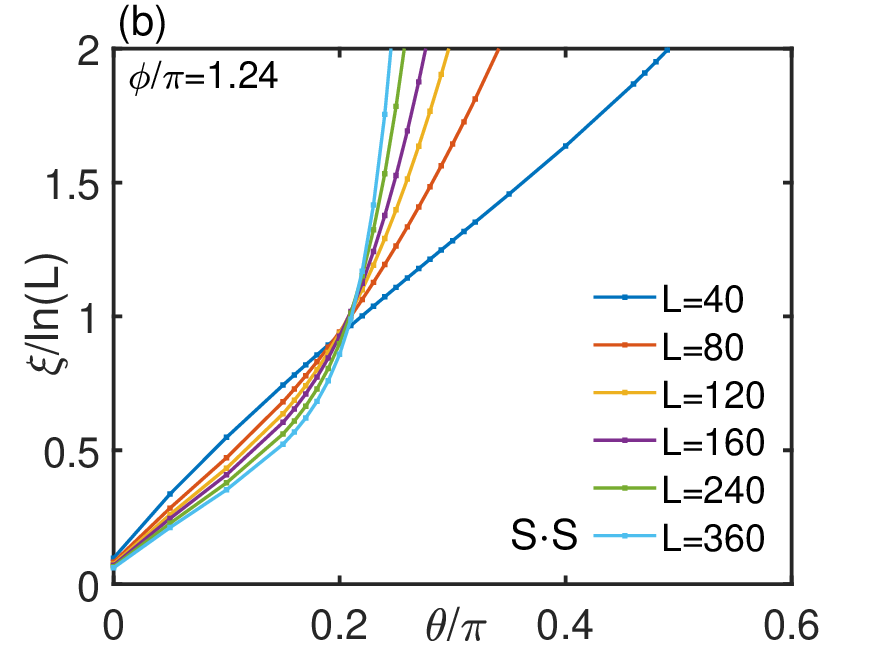}
			%\caption{fig2}
		\end{minipage}
	}%
	\quad
	\subfigure{
		\begin{minipage}{0.5\linewidth}
			\centering
			\includegraphics[scale=0.31]{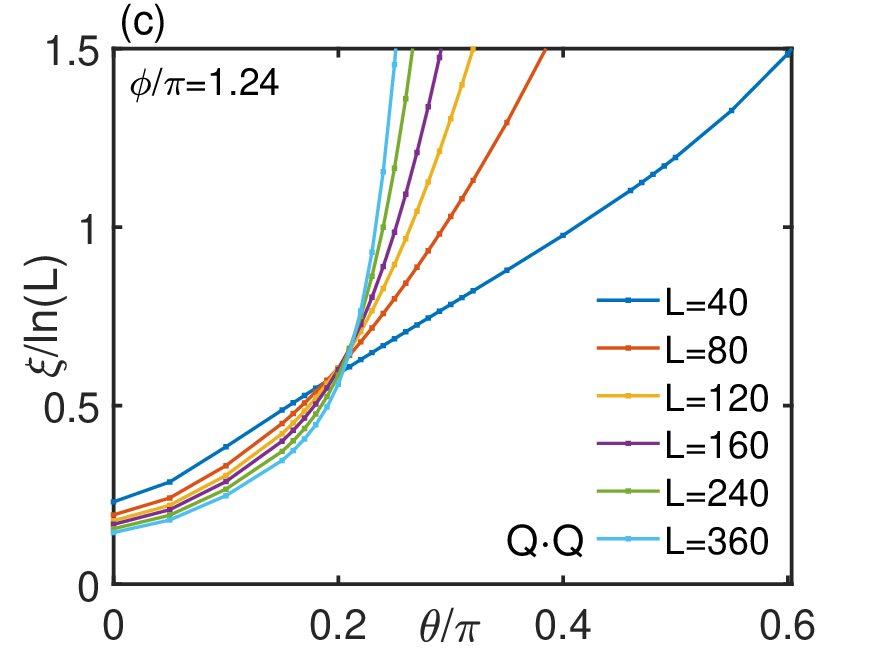}
			%\caption{fig3}
		\end{minipage}
	}%
	\subfigure{
		\begin{minipage}{0.5\linewidth}
			\centering
			\includegraphics[scale=0.31]{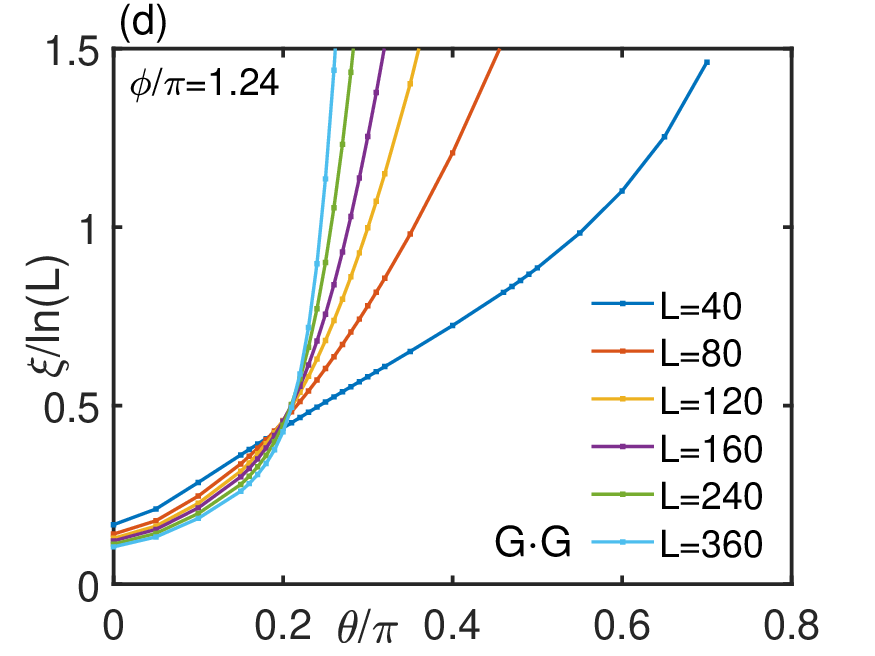}
			%\caption{fig4}
		\end{minipage}
	}%
	
	\caption{Variation of the correlation length $\xi / \ln (L)$ with parameter $\theta$ for different chain lengths of the bond energy calculation obtained for different parts. (a) the Hamiltonian $H$, (b) the spin part $g_1(\mathbf{S} \cdot \mathbf{S})$, (c) the quadrupole operator part $g_2(\mathbf{Q} \cdot \mathbf{Q})$, (d) the octupole operator part $g_3(\mathbf{G} \cdot \mathbf{G} )$. The inset in (a) gives the trend of the effective parameter $\delta$ obtained from these four bond calculations.}
	\label{pic4}
	
\end{figure}

We separately calculate the three bond energies, $\left[ g_1\left( \mathbf{S}\cdot \mathbf{S} \right) ,g_2\left( \mathbf{Q}\cdot \mathbf{Q} \right) ,g_3\left( \mathbf{G}\cdot \mathbf{G} \right) \right] $, and find they all form a similar two-cycle case of the even-odd bonds. We can use each of them to calculate the respective effective parameters $\delta=\left|E_{L / 2}^{b o n d}-E_{L / 2+1}^{b o n d}\right| /\left|E_{L / 2}^{b o n d}+E_{L / 2+1}^{b o n d}\right|$ to obtain the correlation length in Eq. \ref{gap}. As shown in Fig.~\ref{pic4}(b)(c)(d), they all exhibit well criticality and all point to the same phase transition point. As depicted in Fig.~\ref{pic4}(a), the overall correlation length, as determined by the Hamiltonian $H$, manifests in the criticality of the dimer to Luttinger liquid phase transition, and it also drastically changes at the ferromagnetic to the dimer phase transition. As can be seen from the Fig.~\ref{pic4}, the correlation length keeps getting larger in the Luttinger liquid region as the size increases, which implies that the correlation length is infinite in the thermodynamic limit, which is reasonable since the Luttinger liquid phase is critical and gapless. As shown in the inset of the Fig.~\ref{pic4}(a), we do observe the closing of the gap. We obtained the phase transition boundary from dimer to Luttinger liquid using this method, as shown in Fig.~\ref{pic1}. Some theories suggest that the critical theory for general half-integer spin antiferromagnets is the WZW model with topological coupling with level $k = 1$\cite{affleck1987critical,affleck1986exact}; our numerical evidence supports this argument.

As pointed out by Haldane, the one-dimensional antiferromagnetic Heisenberg model is very different when the spin is half-integer and integer, which is due to topological effects\cite{haldane19883,affleck1987critical}. In the generalized Heisenberg model with $S=1$, there is a Haldane phase with the energy gap; and at $S=1/2$ and $3/2$ cases, there will be no Haldane phase; instead, there is Luttinger liquid generated. In the generalized Heisenberg model with $S=1$, there is a phase transition from dimer to Haldane phase, and here there is a phase transition from dimer to Luttinger liquid at $S=3/2$, they have different mechanisms to generate dimer, which leads to different energy gap formulas. Hu's work\cite{Hu14} shows that the dimer in the case of $S=1$ is due to the fluctuations of the Berry phase. At this time the energy gap formula is $\Delta \propto 1 / \xi \approx \exp (-\pi \sqrt{2 / \Delta \theta})$.

%---------------------------------------------------------------------------------------
%---------------------------------------------------------------------------------------

{\it {\color{blue}Summary \& outlook.}}---In the one-dimensional generalized Heisenberg model, we successfully reorganize the Hamiltonian \eqref{Ham} into a sum of multipole bilinear operators for spin-$3/2$.
This allows us to characterize two SU($4$) ULS points effectively.
Through large-scale DMRG simulations, we accurately determine the ground-state phase diagram.
In doing so, we identify two non-uniform spin nematic states arising from the spontaneous translation symmetry breaking in the stripe-Q and tetramer phase regions.
While the dimerization is influenced by bicubic terms, the transition to the Luttinger liquid phase maintains the same universality as the WZW minimal model with level $k = 1$.
This intriguing connection calls for further exploration and investigation.

Further investigation is needed to explore the microscopic mechanisms and physical pictures governing the formation of phases, particularly the non-uniform stripe-Q and tetramer phases.
As the trimer phase in $S=1$ is critical, whether the tetramer phase here is critical needs further analysis.
%------------------------------------------------------------------------------------------------------------------------------------------------------------------------------------------------------------
\bibliography{ref}

\begin{thebibliography}{35}%
\makeatletter
\providecommand \@ifxundefined [1]{%
 \@ifx{#1\undefined}
}%
\providecommand \@ifnum [1]{%
 \ifnum #1\expandafter \@firstoftwo
 \else \expandafter \@secondoftwo
 \fi
}%
\providecommand \@ifx [1]{%
 \ifx #1\expandafter \@firstoftwo
 \else \expandafter \@secondoftwo
 \fi
}%
\providecommand \natexlab [1]{#1}%
\providecommand \enquote  [1]{``#1''}%
\providecommand \bibnamefont  [1]{#1}%
\providecommand \bibfnamefont [1]{#1}%
\providecommand \citenamefont [1]{#1}%
\providecommand \href@noop [0]{\@secondoftwo}%
\providecommand \href [0]{\begingroup \@sanitize@url \@href}%
\providecommand \@href[1]{\@@startlink{#1}\@@href}%
\providecommand \@@href[1]{\endgroup#1\@@endlink}%
\providecommand \@sanitize@url [0]{\catcode `\\12\catcode `\$12\catcode
  `\&12\catcode `\#12\catcode `\^12\catcode `\_12\catcode `\%12\relax}%
\providecommand \@@startlink[1]{}%
\providecommand \@@endlink[0]{}%
\providecommand \url  [0]{\begingroup\@sanitize@url \@url }%
\providecommand \@url [1]{\endgroup\@href {#1}{\urlprefix }}%
\providecommand \urlprefix  [0]{URL }%
\providecommand \Eprint [0]{\href }%
\providecommand \doibase [0]{https://doi.org/}%
\providecommand \selectlanguage [0]{\@gobble}%
\providecommand \bibinfo  [0]{\@secondoftwo}%
\providecommand \bibfield  [0]{\@secondoftwo}%
\providecommand \translation [1]{[#1]}%
\providecommand \BibitemOpen [0]{}%
\providecommand \bibitemStop [0]{}%
\providecommand \bibitemNoStop [0]{.\EOS\space}%
\providecommand \EOS [0]{\spacefactor3000\relax}%
\providecommand \BibitemShut  [1]{\csname bibitem#1\endcsname}%
\let\auto@bib@innerbib\@empty
%</preamble>
\bibitem [{\citenamefont {Soda}\ \emph {et~al.}(2014)\citenamefont {Soda},
  \citenamefont {Matsumoto}, \citenamefont {M{\aa}nsson}, \citenamefont
  {Ohira-Kawamura}, \citenamefont {Nakajima}, \citenamefont {Shiina},\ and\
  \citenamefont {Masuda}}]{Soda_2014}%
  \BibitemOpen
  \bibfield  {author} {\bibinfo {author} {\bibfnamefont {M.}~\bibnamefont
  {Soda}}, \bibinfo {author} {\bibfnamefont {M.}~\bibnamefont {Matsumoto}},
  \bibinfo {author} {\bibfnamefont {M.}~\bibnamefont {M{\aa}nsson}}, \bibinfo
  {author} {\bibfnamefont {S.}~\bibnamefont {Ohira-Kawamura}}, \bibinfo
  {author} {\bibfnamefont {K.}~\bibnamefont {Nakajima}}, \bibinfo {author}
  {\bibfnamefont {R.}~\bibnamefont {Shiina}},\ and\ \bibinfo {author}
  {\bibfnamefont {T.}~\bibnamefont {Masuda}},\ }\bibfield  {title} {\bibinfo
  {title} {Spin-nematic interaction in the multiferroic compound ba 2 coge 2 o
  7},\ }\href {https://doi.org/10.1103%2Fphysrevlett.112.127205} {\bibfield
  {journal} {\bibinfo  {journal} {Phys. Rev. Lett.}\ }\textbf {\bibinfo
  {volume} {112}},\ \bibinfo {pages} {127205} (\bibinfo {year}
  {2014})}\BibitemShut {NoStop}%
\bibitem [{\citenamefont {Watanabe}\ \emph {et~al.}(2023)\citenamefont
  {Watanabe}, \citenamefont {Miyake}, \citenamefont {Gen}, \citenamefont
  {Mizukami}, \citenamefont {Hashimoto}, \citenamefont {Shibauchi},
  \citenamefont {Ikeda}, \citenamefont {Tokunaga}, \citenamefont {Kurumaji},
  \citenamefont {Tokunaga} \emph {et~al.}}]{Watanabe_2023}%
  \BibitemOpen
  \bibfield  {author} {\bibinfo {author} {\bibfnamefont {Y.}~\bibnamefont
  {Watanabe}}, \bibinfo {author} {\bibfnamefont {A.}~\bibnamefont {Miyake}},
  \bibinfo {author} {\bibfnamefont {M.}~\bibnamefont {Gen}}, \bibinfo {author}
  {\bibfnamefont {Y.}~\bibnamefont {Mizukami}}, \bibinfo {author}
  {\bibfnamefont {K.}~\bibnamefont {Hashimoto}}, \bibinfo {author}
  {\bibfnamefont {T.}~\bibnamefont {Shibauchi}}, \bibinfo {author}
  {\bibfnamefont {A.}~\bibnamefont {Ikeda}}, \bibinfo {author} {\bibfnamefont
  {M.}~\bibnamefont {Tokunaga}}, \bibinfo {author} {\bibfnamefont
  {T.}~\bibnamefont {Kurumaji}}, \bibinfo {author} {\bibfnamefont
  {Y.}~\bibnamefont {Tokunaga}}, \emph {et~al.},\ }\bibfield  {title} {\bibinfo
  {title} {Double dome structure of the bose--einstein condensation in diluted
  s= 3/2 quantum magnets},\ }\href
  {https://doi.org/10.1038%2Fs41467-023-36725-4} {\bibfield  {journal}
  {\bibinfo  {journal} {Nat. Commun.}\ }\textbf {\bibinfo {volume} {14}},\
  \bibinfo {pages} {1260} (\bibinfo {year} {2023})}\BibitemShut {NoStop}%
\bibitem [{\citenamefont {Brydon}\ \emph {et~al.}(2016)\citenamefont {Brydon},
  \citenamefont {Wang}, \citenamefont {Weinert},\ and\ \citenamefont
  {Agterberg}}]{Brydon_2016}%
  \BibitemOpen
  \bibfield  {author} {\bibinfo {author} {\bibfnamefont {P.}~\bibnamefont
  {Brydon}}, \bibinfo {author} {\bibfnamefont {L.}~\bibnamefont {Wang}},
  \bibinfo {author} {\bibfnamefont {M.}~\bibnamefont {Weinert}},\ and\ \bibinfo
  {author} {\bibfnamefont {D.}~\bibnamefont {Agterberg}},\ }\bibfield  {title}
  {\bibinfo {title} {Pairing of j= 3/2 fermions in half-heusler
  superconductors},\ }\href {https://doi.org/10.1103%2Fphysrevlett.116.177001}
  {\bibfield  {journal} {\bibinfo  {journal} {Phys. Rev. Lett.}\ }\textbf
  {\bibinfo {volume} {116}},\ \bibinfo {pages} {177001} (\bibinfo {year}
  {2016})}\BibitemShut {NoStop}%
\bibitem [{\citenamefont {Kim}\ \emph {et~al.}(2018)\citenamefont {Kim},
  \citenamefont {Wang}, \citenamefont {Nakajima}, \citenamefont {Hu},
  \citenamefont {Ziemak}, \citenamefont {Syers}, \citenamefont {Wang},
  \citenamefont {Hodovanets}, \citenamefont {Denlinger}, \citenamefont {Brydon}
  \emph {et~al.}}]{Kim_2018}%
  \BibitemOpen
  \bibfield  {author} {\bibinfo {author} {\bibfnamefont {H.}~\bibnamefont
  {Kim}}, \bibinfo {author} {\bibfnamefont {K.}~\bibnamefont {Wang}}, \bibinfo
  {author} {\bibfnamefont {Y.}~\bibnamefont {Nakajima}}, \bibinfo {author}
  {\bibfnamefont {R.}~\bibnamefont {Hu}}, \bibinfo {author} {\bibfnamefont
  {S.}~\bibnamefont {Ziemak}}, \bibinfo {author} {\bibfnamefont
  {P.}~\bibnamefont {Syers}}, \bibinfo {author} {\bibfnamefont
  {L.}~\bibnamefont {Wang}}, \bibinfo {author} {\bibfnamefont {H.}~\bibnamefont
  {Hodovanets}}, \bibinfo {author} {\bibfnamefont {J.~D.}\ \bibnamefont
  {Denlinger}}, \bibinfo {author} {\bibfnamefont {P.~M.}\ \bibnamefont
  {Brydon}}, \emph {et~al.},\ }\bibfield  {title} {\bibinfo {title} {Beyond
  triplet: Unconventional superconductivity in a spin-3/2 topological
  semimetal},\ }\href {https://doi.org/10.1126%2Fsciadv.aao4513} {\bibfield
  {journal} {\bibinfo  {journal} {Sci. Adv.}\ }\textbf {\bibinfo {volume}
  {4}},\ \bibinfo {pages} {eaao4513} (\bibinfo {year} {2018})}\BibitemShut
  {NoStop}%
\bibitem [{\citenamefont {Zhang}\ \emph {et~al.}(2021)\citenamefont {Zhang},
  \citenamefont {Wu}, \citenamefont {Kasahara}, \citenamefont {Shibauchi},
  \citenamefont {Matsuda},\ and\ \citenamefont {Blumberg}}]{Zhang_2021}%
  \BibitemOpen
  \bibfield  {author} {\bibinfo {author} {\bibfnamefont {W.}~\bibnamefont
  {Zhang}}, \bibinfo {author} {\bibfnamefont {S.}~\bibnamefont {Wu}}, \bibinfo
  {author} {\bibfnamefont {S.}~\bibnamefont {Kasahara}}, \bibinfo {author}
  {\bibfnamefont {T.}~\bibnamefont {Shibauchi}}, \bibinfo {author}
  {\bibfnamefont {Y.}~\bibnamefont {Matsuda}},\ and\ \bibinfo {author}
  {\bibfnamefont {G.}~\bibnamefont {Blumberg}},\ }\bibfield  {title} {\bibinfo
  {title} {Quadrupolar charge dynamics in the nonmagnetic fese1- x s x
  superconductors},\ }\href {https://doi.org/10.1073%2Fpnas.2020585118}
  {\bibfield  {journal} {\bibinfo  {journal} {Proc. Nat. Acad. Sci.}\ }\textbf
  {\bibinfo {volume} {118}},\ \bibinfo {pages} {e2020585118} (\bibinfo {year}
  {2021})}\BibitemShut {NoStop}%
\bibitem [{\citenamefont {Xu}\ \emph {et~al.}(2020)\citenamefont {Xu},
  \citenamefont {Feng}, \citenamefont {Kawamura}, \citenamefont {Yamaji},
  \citenamefont {Nahas}, \citenamefont {Prokhorenko}, \citenamefont {Qi},
  \citenamefont {Xiang},\ and\ \citenamefont {Bellaiche}}]{Xu_2020}%
  \BibitemOpen
  \bibfield  {author} {\bibinfo {author} {\bibfnamefont {C.}~\bibnamefont
  {Xu}}, \bibinfo {author} {\bibfnamefont {J.}~\bibnamefont {Feng}}, \bibinfo
  {author} {\bibfnamefont {M.}~\bibnamefont {Kawamura}}, \bibinfo {author}
  {\bibfnamefont {Y.}~\bibnamefont {Yamaji}}, \bibinfo {author} {\bibfnamefont
  {Y.}~\bibnamefont {Nahas}}, \bibinfo {author} {\bibfnamefont
  {S.}~\bibnamefont {Prokhorenko}}, \bibinfo {author} {\bibfnamefont
  {Y.}~\bibnamefont {Qi}}, \bibinfo {author} {\bibfnamefont {H.}~\bibnamefont
  {Xiang}},\ and\ \bibinfo {author} {\bibfnamefont {L.}~\bibnamefont
  {Bellaiche}},\ }\bibfield  {title} {\bibinfo {title} {Possible kitaev quantum
  spin liquid state in 2d materials with s= 3/2},\ }\href
  {https://doi.org/10.1103%2Fphysrevlett.124.087205} {\bibfield  {journal}
  {\bibinfo  {journal} {Phys. Rev. Lett.}\ }\textbf {\bibinfo {volume} {124}},\
  \bibinfo {pages} {087205} (\bibinfo {year} {2020})}\BibitemShut {NoStop}%
\bibitem [{\citenamefont {Hallberg}\ \emph {et~al.}(1996)\citenamefont
  {Hallberg}, \citenamefont {Wang}, \citenamefont {Horsch},\ and\ \citenamefont
  {Moreo}}]{Hallberg_1996}%
  \BibitemOpen
  \bibfield  {author} {\bibinfo {author} {\bibfnamefont {K.}~\bibnamefont
  {Hallberg}}, \bibinfo {author} {\bibfnamefont {X.}~\bibnamefont {Wang}},
  \bibinfo {author} {\bibfnamefont {P.}~\bibnamefont {Horsch}},\ and\ \bibinfo
  {author} {\bibfnamefont {A.}~\bibnamefont {Moreo}},\ }\bibfield  {title}
  {\bibinfo {title} {Critical behavior of the s= 3/2 antiferromagnetic
  heisenberg chain},\ }\href {https://doi.org/10.1103%2Fphysrevlett.76.4955}
  {\bibfield  {journal} {\bibinfo  {journal} {Phys. Rev. Lett.}\ }\textbf
  {\bibinfo {volume} {76}},\ \bibinfo {pages} {4955} (\bibinfo {year}
  {1996})}\BibitemShut {NoStop}%
\bibitem [{\citenamefont {Haldane}(1988)}]{haldane19883}%
  \BibitemOpen
  \bibfield  {author} {\bibinfo {author} {\bibfnamefont {F.~D.~M.}\
  \bibnamefont {Haldane}},\ }\bibfield  {title} {\bibinfo {title} {O(3)
  nonlinear $\sigma$ model and the topological distinction between integer-and
  half-integer-spin antiferromagnets in two dimensions},\ }\href
  {https://journals.aps.org/prl/abstract/10.1103/PhysRevLett.61.1029}
  {\bibfield  {journal} {\bibinfo  {journal} {Phys. Rev. Lett.}\ }\textbf
  {\bibinfo {volume} {61}},\ \bibinfo {pages} {1029} (\bibinfo {year}
  {1988})}\BibitemShut {NoStop}%
\bibitem [{\citenamefont {Kugel'}\ and\ \citenamefont
  {Khomski{\u{\i}}}(1982)}]{Kugel_1982}%
  \BibitemOpen
  \bibfield  {author} {\bibinfo {author} {\bibfnamefont {K.~I.}\ \bibnamefont
  {Kugel'}}\ and\ \bibinfo {author} {\bibfnamefont {D.~I.}\ \bibnamefont
  {Khomski{\u{\i}}}},\ }\bibfield  {title} {\bibinfo {title} {The jahn-teller
  effect and magnetism: transition metal compounds},\ }\href
  {https://doi.org/10.1070/pu1982v025n04abeh004537} {\bibfield  {journal}
  {\bibinfo  {journal} {Sov. phys. Usp.}\ }\textbf {\bibinfo {volume} {25}},\
  \bibinfo {pages} {231} (\bibinfo {year} {1982})}\BibitemShut {NoStop}%
\bibitem [{\citenamefont {Daghofer}\ \emph {et~al.}(2008)\citenamefont
  {Daghofer}, \citenamefont {Wohlfeld}, \citenamefont {Ole{\'s}}, \citenamefont
  {Arrigoni},\ and\ \citenamefont {Horsch}}]{daghofer2008absence}%
  \BibitemOpen
  \bibfield  {author} {\bibinfo {author} {\bibfnamefont {M.}~\bibnamefont
  {Daghofer}}, \bibinfo {author} {\bibfnamefont {K.}~\bibnamefont {Wohlfeld}},
  \bibinfo {author} {\bibfnamefont {A.~M.}\ \bibnamefont {Ole{\'s}}}, \bibinfo
  {author} {\bibfnamefont {E.}~\bibnamefont {Arrigoni}},\ and\ \bibinfo
  {author} {\bibfnamefont {P.}~\bibnamefont {Horsch}},\ }\bibfield  {title}
  {\bibinfo {title} {Absence of hole confinement in transition-metal oxides
  with orbital degeneracy},\ }\href
  {https://doi.org/https://doi.org/10.1103/PhysRevLett.100.066403} {\bibfield
  {journal} {\bibinfo  {journal} {Phys. Rev. Lett.}\ }\textbf {\bibinfo
  {volume} {100}},\ \bibinfo {pages} {066403} (\bibinfo {year}
  {2008})}\BibitemShut {NoStop}%
\bibitem [{\citenamefont {Wu}\ \emph {et~al.}(2018)\citenamefont {Wu},
  \citenamefont {Lovorn}, \citenamefont {Tutuc},\ and\ \citenamefont
  {MacDonald}}]{wu2018hubbard}%
  \BibitemOpen
  \bibfield  {author} {\bibinfo {author} {\bibfnamefont {F.}~\bibnamefont
  {Wu}}, \bibinfo {author} {\bibfnamefont {T.}~\bibnamefont {Lovorn}}, \bibinfo
  {author} {\bibfnamefont {E.}~\bibnamefont {Tutuc}},\ and\ \bibinfo {author}
  {\bibfnamefont {A.~H.}\ \bibnamefont {MacDonald}},\ }\bibfield  {title}
  {\bibinfo {title} {Hubbard model physics in transition metal dichalcogenide
  moir{\'e} bands},\ }\href
  {https://doi.org/https://doi.org/10.1103/PhysRevLett.121.026402} {\bibfield
  {journal} {\bibinfo  {journal} {Phys. Rev. Lett.}\ }\textbf {\bibinfo
  {volume} {121}},\ \bibinfo {pages} {026402} (\bibinfo {year}
  {2018})}\BibitemShut {NoStop}%
\bibitem [{\citenamefont {Janani}\ \emph {et~al.}(2014)\citenamefont {Janani},
  \citenamefont {Merino}, \citenamefont {McCulloch},\ and\ \citenamefont
  {Powell}}]{janani2014haldane}%
  \BibitemOpen
  \bibfield  {author} {\bibinfo {author} {\bibfnamefont {C.}~\bibnamefont
  {Janani}}, \bibinfo {author} {\bibfnamefont {J.}~\bibnamefont {Merino}},
  \bibinfo {author} {\bibfnamefont {I.}~\bibnamefont {McCulloch}},\ and\
  \bibinfo {author} {\bibfnamefont {B.}~\bibnamefont {Powell}},\ }\bibfield
  {title} {\bibinfo {title} {Haldane phase in the hubbard model at 2/3-filling
  for the organic molecular compound mo 3 s 7 (dmit) 3},\ }\href
  {https://doi.org/https://doi.org/10.1103/PhysRevLett.113.267204} {\bibfield
  {journal} {\bibinfo  {journal} {Phys. Rev. Lett.}\ }\textbf {\bibinfo
  {volume} {113}},\ \bibinfo {pages} {267204} (\bibinfo {year}
  {2014})}\BibitemShut {NoStop}%
\bibitem [{\citenamefont {Dagotto}\ \emph {et~al.}(2011)\citenamefont
  {Dagotto}, \citenamefont {Moreo}, \citenamefont {Nicholson}, \citenamefont
  {Luo}, \citenamefont {Liang},\ and\ \citenamefont
  {Zhang}}]{dagotto2011properties}%
  \BibitemOpen
  \bibfield  {author} {\bibinfo {author} {\bibfnamefont {E.}~\bibnamefont
  {Dagotto}}, \bibinfo {author} {\bibfnamefont {A.}~\bibnamefont {Moreo}},
  \bibinfo {author} {\bibfnamefont {A.}~\bibnamefont {Nicholson}}, \bibinfo
  {author} {\bibfnamefont {Q.}~\bibnamefont {Luo}}, \bibinfo {author}
  {\bibfnamefont {S.}~\bibnamefont {Liang}},\ and\ \bibinfo {author}
  {\bibfnamefont {X.}~\bibnamefont {Zhang}},\ }\bibfield  {title} {\bibinfo
  {title} {Properties of the multiorbital hubbard models for the iron-based
  superconductors},\ }\href
  {https://link.springer.com/article/10.1007/s11467-011-0222-z} {\bibfield
  {journal} {\bibinfo  {journal} {Front Phys}\ }\textbf {\bibinfo {volume}
  {6}},\ \bibinfo {pages} {379} (\bibinfo {year} {2011})}\BibitemShut {NoStop}%
\bibitem [{\citenamefont {Takimoto}\ \emph {et~al.}(2004)\citenamefont
  {Takimoto}, \citenamefont {Hotta},\ and\ \citenamefont
  {Ueda}}]{takimoto2004strong}%
  \BibitemOpen
  \bibfield  {author} {\bibinfo {author} {\bibfnamefont {T.}~\bibnamefont
  {Takimoto}}, \bibinfo {author} {\bibfnamefont {T.}~\bibnamefont {Hotta}},\
  and\ \bibinfo {author} {\bibfnamefont {K.}~\bibnamefont {Ueda}},\ }\bibfield
  {title} {\bibinfo {title} {Strong-coupling theory of superconductivity in a
  degenerate hubbard model},\ }\href
  {https://doi.org/https://doi.org/10.1103/PhysRevB.69.104504} {\bibfield
  {journal} {\bibinfo  {journal} {Phys. Rev. B}\ }\textbf {\bibinfo {volume}
  {69}},\ \bibinfo {pages} {104504} (\bibinfo {year} {2004})}\BibitemShut
  {NoStop}%
\bibitem [{\citenamefont {Dutta}\ \emph {et~al.}(2015)\citenamefont {Dutta},
  \citenamefont {Gajda}, \citenamefont {Hauke}, \citenamefont {Lewenstein},
  \citenamefont {L{\"u}hmann}, \citenamefont {Malomed}, \citenamefont
  {Sowi{\'n}ski},\ and\ \citenamefont {Zakrzewski}}]{dutta2015non}%
  \BibitemOpen
  \bibfield  {author} {\bibinfo {author} {\bibfnamefont {O.}~\bibnamefont
  {Dutta}}, \bibinfo {author} {\bibfnamefont {M.}~\bibnamefont {Gajda}},
  \bibinfo {author} {\bibfnamefont {P.}~\bibnamefont {Hauke}}, \bibinfo
  {author} {\bibfnamefont {M.}~\bibnamefont {Lewenstein}}, \bibinfo {author}
  {\bibfnamefont {D.-S.}\ \bibnamefont {L{\"u}hmann}}, \bibinfo {author}
  {\bibfnamefont {B.~A.}\ \bibnamefont {Malomed}}, \bibinfo {author}
  {\bibfnamefont {T.}~\bibnamefont {Sowi{\'n}ski}},\ and\ \bibinfo {author}
  {\bibfnamefont {J.}~\bibnamefont {Zakrzewski}},\ }\bibfield  {title}
  {\bibinfo {title} {Non-standard hubbard models in optical lattices: a
  review},\ }\href
  {https://iopscience.iop.org/article/10.1088/0034-4885/78/6/066001/meta}
  {\bibfield  {journal} {\bibinfo  {journal} {Rep. Prog. Phys.}\ }\textbf
  {\bibinfo {volume} {78}},\ \bibinfo {pages} {066001} (\bibinfo {year}
  {2015})}\BibitemShut {NoStop}%
\bibitem [{\citenamefont {Yamashita}\ \emph {et~al.}(1998)\citenamefont
  {Yamashita}, \citenamefont {Shibata},\ and\ \citenamefont
  {Ueda}}]{Yamashita98}%
  \BibitemOpen
  \bibfield  {author} {\bibinfo {author} {\bibfnamefont {Y.}~\bibnamefont
  {Yamashita}}, \bibinfo {author} {\bibfnamefont {N.}~\bibnamefont {Shibata}},\
  and\ \bibinfo {author} {\bibfnamefont {K.}~\bibnamefont {Ueda}},\ }\bibfield
  {title} {\bibinfo {title} {{$SU(4)$ spin-orbit critical state in one
  dimension}},\ }\href
  {https://doi.org/https://doi.org/10.1103/PhysRevB.58.9114} {\bibfield
  {journal} {\bibinfo  {journal} {Phys. Rev. B}\ }\textbf {\bibinfo {volume}
  {58}},\ \bibinfo {pages} {9114} (\bibinfo {year} {1998})}\BibitemShut
  {NoStop}%
\bibitem [{\citenamefont {Fridman}\ \emph {et~al.}(2011)\citenamefont
  {Fridman}, \citenamefont {Kosmachev}, \citenamefont {Kolezhuk},\ and\
  \citenamefont {Ivanov}}]{Fridman_2011}%
  \BibitemOpen
  \bibfield  {author} {\bibinfo {author} {\bibfnamefont {Y.~A.}\ \bibnamefont
  {Fridman}}, \bibinfo {author} {\bibfnamefont {O.}~\bibnamefont {Kosmachev}},
  \bibinfo {author} {\bibfnamefont {A.}~\bibnamefont {Kolezhuk}},\ and\
  \bibinfo {author} {\bibfnamefont {B.}~\bibnamefont {Ivanov}},\ }\bibfield
  {title} {\bibinfo {title} {Spin nematic and antinematic states in a spin-3 2
  isotropic non-heisenberg magnet},\ }\href
  {https://doi.org/10.1103%2Fphysrevlett.106.097202} {\bibfield  {journal}
  {\bibinfo  {journal} {Phys. Rev. Lett.}\ }\textbf {\bibinfo {volume} {106}},\
  \bibinfo {pages} {097202} (\bibinfo {year} {2011})}\BibitemShut {NoStop}%
\bibitem [{\citenamefont {Fáth}\ and\ \citenamefont {Sólyom}(1991)}]{Fath91}%
  \BibitemOpen
  \bibfield  {author} {\bibinfo {author} {\bibfnamefont {G.}~\bibnamefont
  {Fáth}}\ and\ \bibinfo {author} {\bibfnamefont {J.}~\bibnamefont
  {Sólyom}},\ }\bibfield  {title} {\bibinfo {title} {{Period tripling in the
  bilinear-biquadratic antiferromagnetic S=1 chain}},\ }\href
  {https://doi.org/https://doi.org/10.1103/PhysRevB.44.11836} {\bibfield
  {journal} {\bibinfo  {journal} {Phys. Rev. B}\ }\textbf {\bibinfo {volume}
  {44}},\ \bibinfo {pages} {11836} (\bibinfo {year} {1991})}\BibitemShut
  {NoStop}%
\bibitem [{\citenamefont {Fáth}\ and\ \citenamefont {Sólyom}(1995)}]{Fath95}%
  \BibitemOpen
  \bibfield  {author} {\bibinfo {author} {\bibfnamefont {G.}~\bibnamefont
  {Fáth}}\ and\ \bibinfo {author} {\bibfnamefont {J.}~\bibnamefont
  {Sólyom}},\ }\bibfield  {title} {\bibinfo {title} {{Search for the
  nondimerized quantum nematic phase in the spin-1 chain}},\ }\href
  {https://doi.org/https://doi.org/10.1103/PhysRevB.51.3620} {\bibfield
  {journal} {\bibinfo  {journal} {Phys. Rev. B}\ }\textbf {\bibinfo {volume}
  {51}},\ \bibinfo {pages} {3620} (\bibinfo {year} {1995})}\BibitemShut
  {NoStop}%
\bibitem [{\citenamefont {Hu}\ \emph {et~al.}(2014)\citenamefont {Hu},
  \citenamefont {Turner}, \citenamefont {Penc},\ and\ \citenamefont
  {Pollmann}}]{Hu14}%
  \BibitemOpen
  \bibfield  {author} {\bibinfo {author} {\bibfnamefont {S.}~\bibnamefont
  {Hu}}, \bibinfo {author} {\bibfnamefont {A.~M.}\ \bibnamefont {Turner}},
  \bibinfo {author} {\bibfnamefont {K.}~\bibnamefont {Penc}},\ and\ \bibinfo
  {author} {\bibfnamefont {F.}~\bibnamefont {Pollmann}},\ }\bibfield  {title}
  {\bibinfo {title} {{Berry-Phase-Induced Dimerization in One-Dimensional
  Quadrupolar Systems}},\ }\href
  {https://doi.org/https://doi.org/10.1103/PhysRevLett.113.027202} {\bibfield
  {journal} {\bibinfo  {journal} {Phys. Rev. Lett.}\ }\textbf {\bibinfo
  {volume} {113}},\ \bibinfo {pages} {027202} (\bibinfo {year}
  {2014})}\BibitemShut {NoStop}%
\bibitem [{\citenamefont {Itoi}\ and\ \citenamefont {Kato}(1997)}]{Itoi97}%
  \BibitemOpen
  \bibfield  {author} {\bibinfo {author} {\bibfnamefont {C.}~\bibnamefont
  {Itoi}}\ and\ \bibinfo {author} {\bibfnamefont {M.-H.}\ \bibnamefont
  {Kato}},\ }\bibfield  {title} {\bibinfo {title} {{Extended massless phase and
  the Haldane phase in a spin-1 isotropic antiferromagnetic chain}},\ }\href
  {https://doi.org/10.1103/PhysRevB.55.8295} {\bibfield  {journal} {\bibinfo
  {journal} {Phys. Rev. B}\ }\textbf {\bibinfo {volume} {55}},\ \bibinfo
  {pages} {8295} (\bibinfo {year} {1997})}\BibitemShut {NoStop}%
\bibitem [{\citenamefont {Uimin}(1970)}]{Uimin70}%
  \BibitemOpen
  \bibfield  {author} {\bibinfo {author} {\bibfnamefont {G.~V.}\ \bibnamefont
  {Uimin}},\ }\bibfield  {title} {\bibinfo {title} {{One-dimensional Problem
  for S = 1 with Modified Antiferromagnetic Hamiltonian}},\ }\href
  {https://ui.adsabs.harvard.edu/abs/1970JETPL..12..225U/abstract} {\bibfield
  {journal} {\bibinfo  {journal} {JETP Lett.}\ }\textbf {\bibinfo {volume}
  {12}},\ \bibinfo {pages} {332} (\bibinfo {year} {1970})}\BibitemShut
  {NoStop}%
\bibitem [{\citenamefont {Lai}(1974)}]{Lai74}%
  \BibitemOpen
  \bibfield  {author} {\bibinfo {author} {\bibfnamefont {C.~K.}\ \bibnamefont
  {Lai}},\ }\bibfield  {title} {\bibinfo {title} {{Lattice gas with
  nearest‐neighbor interaction in one dimension with arbitrary statistics}},\
  }\href {https://aip.scitation.org/doi/10.1063/1.1666522} {\bibfield
  {journal} {\bibinfo  {journal} {J. Math. Phys.}\ }\textbf {\bibinfo {volume}
  {15}},\ \bibinfo {pages} {1675} (\bibinfo {year} {1974})}\BibitemShut
  {NoStop}%
\bibitem [{\citenamefont {Sutherland}(1975)}]{Sutherland75}%
  \BibitemOpen
  \bibfield  {author} {\bibinfo {author} {\bibfnamefont {B.}~\bibnamefont
  {Sutherland}},\ }\bibfield  {title} {\bibinfo {title} {{Model for a
  multicomponent quantum system}},\ }\href
  {https://doi.org/https://doi.org/10.1103/PhysRevB.12.3795} {\bibfield
  {journal} {\bibinfo  {journal} {Phys. Rev. B}\ }\textbf {\bibinfo {volume}
  {12}},\ \bibinfo {pages} {3795} (\bibinfo {year} {1975})}\BibitemShut
  {NoStop}%
\bibitem [{\citenamefont {Affleck}\ \emph {et~al.}(1987)\citenamefont
  {Affleck}, \citenamefont {Kennedy}, \citenamefont {Lieb},\ and\ \citenamefont
  {Tasaki}}]{AKLT87}%
  \BibitemOpen
  \bibfield  {author} {\bibinfo {author} {\bibfnamefont {I.}~\bibnamefont
  {Affleck}}, \bibinfo {author} {\bibfnamefont {T.}~\bibnamefont {Kennedy}},
  \bibinfo {author} {\bibfnamefont {E.~H.}\ \bibnamefont {Lieb}},\ and\
  \bibinfo {author} {\bibfnamefont {H.}~\bibnamefont {Tasaki}},\ }\bibfield
  {title} {\bibinfo {title} {{Rigorous results on valence-bond ground states in
  antiferromagnets}},\ }\href
  {https://doi.org/https://doi.org/10.1103/PhysRevLett.59.799} {\bibfield
  {journal} {\bibinfo  {journal} {Phys. Rev. Lett.}\ }\textbf {\bibinfo
  {volume} {59}},\ \bibinfo {pages} {799} (\bibinfo {year} {1987})}\BibitemShut
  {NoStop}%
\bibitem [{\citenamefont {Takhtajan}(1982)}]{Takhtajan82}%
  \BibitemOpen
  \bibfield  {author} {\bibinfo {author} {\bibfnamefont {L.}~\bibnamefont
  {Takhtajan}},\ }\bibfield  {title} {\bibinfo {title} {{The picture of
  low-lying excitations in the isotropic Heisenberg chain of arbitrary
  spins}},\ }\href
  {https://doi.org/https://doi.org/10.1016/0375-9601(82)90764-2} {\bibfield
  {journal} {\bibinfo  {journal} {Phys. Lett. A.}\ }\textbf {\bibinfo {volume}
  {87}},\ \bibinfo {pages} {9, 479–482} (\bibinfo {year} {1982})}\BibitemShut
  {NoStop}%
\bibitem [{\citenamefont {Babujian}(1982)}]{Babujian82}%
  \BibitemOpen
  \bibfield  {author} {\bibinfo {author} {\bibfnamefont {H.}~\bibnamefont
  {Babujian}},\ }\bibfield  {title} {\bibinfo {title} {{Exact solution of the
  one-dimensional isotropic Heisenberg chain with arbitrary spins S}},\ }\href
  {https://doi.org/https://doi.org/10.1016/0375-9601(82)90403-0} {\bibfield
  {journal} {\bibinfo  {journal} {Phys. Lett. A.}\ }\textbf {\bibinfo {volume}
  {90}},\ \bibinfo {pages} {9, 479–482} (\bibinfo {year} {1982})}\BibitemShut
  {NoStop}%
\bibitem [{Sup()}]{SuppMat}%
  \BibitemOpen
  \href@noop {} {}\bibinfo {note} {See the Supplementary Material for the
  definition of spherical tensor operators, the proof of several equations, and
  the solution of $SU(4)$ high symmetry point.}\BibitemShut {Stop}%
\bibitem [{\citenamefont {Sandvik}(2010)}]{sandvik2010computational}%
  \BibitemOpen
  \bibfield  {author} {\bibinfo {author} {\bibfnamefont {A.~W.}\ \bibnamefont
  {Sandvik}},\ }\bibfield  {title} {\bibinfo {title} {Computational studies of
  quantum spin systems},\ }in\ \href
  {https://doi.org/https://doi.org/10.1063/1.3518900} {\emph {\bibinfo
  {booktitle} {AIP Conference Proceedings}}},\ Vol.\ \bibinfo {volume} {1297}\
  (\bibinfo {organization} {American Institute of Physics},\ \bibinfo {year}
  {2010})\ pp.\ \bibinfo {pages} {135--338}\BibitemShut {NoStop}%
\bibitem [{\citenamefont {Schollw{\"o}ck}(2011)}]{schollwock2011density}%
  \BibitemOpen
  \bibfield  {author} {\bibinfo {author} {\bibfnamefont {U.}~\bibnamefont
  {Schollw{\"o}ck}},\ }\bibfield  {title} {\bibinfo {title} {The density-matrix
  renormalization group in the age of matrix product states},\ }\href
  {https://www.sciencedirect.com/science/article/abs/pii/S0003491610001752}
  {\bibfield  {journal} {\bibinfo  {journal} {Annals of physics}\ }\textbf
  {\bibinfo {volume} {326}},\ \bibinfo {pages} {96} (\bibinfo {year}
  {2011})}\BibitemShut {NoStop}%
\bibitem [{\citenamefont {Affleck}\ \emph {et~al.}(1989)\citenamefont
  {Affleck}, \citenamefont {Gepner}, \citenamefont {Schulz},\ and\
  \citenamefont {Ziman}}]{affleck1989critical}%
  \BibitemOpen
  \bibfield  {author} {\bibinfo {author} {\bibfnamefont {I.}~\bibnamefont
  {Affleck}}, \bibinfo {author} {\bibfnamefont {D.}~\bibnamefont {Gepner}},
  \bibinfo {author} {\bibfnamefont {H.}~\bibnamefont {Schulz}},\ and\ \bibinfo
  {author} {\bibfnamefont {T.}~\bibnamefont {Ziman}},\ }\bibfield  {title}
  {\bibinfo {title} {Critical behaviour of spin-s heisenberg antiferromagnetic
  chains: analytic and numerical results},\ }\href
  {https://iopscience.iop.org/article/10.1088/0305-4470/22/5/015/meta}
  {\bibfield  {journal} {\bibinfo  {journal} {J. Phys. A: Math. Gen.}\ }\textbf
  {\bibinfo {volume} {22}},\ \bibinfo {pages} {511} (\bibinfo {year}
  {1989})}\BibitemShut {NoStop}%
\bibitem [{\citenamefont {Cardy}(1996)}]{cardy1996scaling}%
  \BibitemOpen
  \bibfield  {author} {\bibinfo {author} {\bibfnamefont {J.}~\bibnamefont
  {Cardy}},\ }\href@noop {} {\emph {\bibinfo {title} {Scaling and
  renormalization in statistical physics}}},\ Vol.~\bibinfo {volume} {5}\
  (\bibinfo  {publisher} {Cambridge university press},\ \bibinfo {year}
  {1996})\BibitemShut {NoStop}%
\bibitem [{\citenamefont {Cardy}(1986)}]{cardy1986logarithmic}%
  \BibitemOpen
  \bibfield  {author} {\bibinfo {author} {\bibfnamefont {J.~L.}\ \bibnamefont
  {Cardy}},\ }\bibfield  {title} {\bibinfo {title} {Logarithmic corrections to
  finite-size scaling in strips},\ }\href
  {https://iopscience.iop.org/article/10.1088/0305-4470/19/17/008} {\bibfield
  {journal} {\bibinfo  {journal} {J. Phys. A: Math. Gen.}\ }\textbf {\bibinfo
  {volume} {19}},\ \bibinfo {pages} {L1093} (\bibinfo {year}
  {1986})}\BibitemShut {NoStop}%
\bibitem [{\citenamefont {Affleck}\ and\ \citenamefont
  {Haldane}(1987)}]{affleck1987critical}%
  \BibitemOpen
  \bibfield  {author} {\bibinfo {author} {\bibfnamefont {I.}~\bibnamefont
  {Affleck}}\ and\ \bibinfo {author} {\bibfnamefont {F.}~\bibnamefont
  {Haldane}},\ }\bibfield  {title} {\bibinfo {title} {Critical theory of
  quantum spin chains},\ }\href
  {https://journals.aps.org/prb/abstract/10.1103/PhysRevB.36.5291} {\bibfield
  {journal} {\bibinfo  {journal} {Phys. Rev. B}\ }\textbf {\bibinfo {volume}
  {36}},\ \bibinfo {pages} {5291} (\bibinfo {year} {1987})}\BibitemShut
  {NoStop}%
\bibitem [{\citenamefont {Affleck}(1986)}]{affleck1986exact}%
  \BibitemOpen
  \bibfield  {author} {\bibinfo {author} {\bibfnamefont {I.}~\bibnamefont
  {Affleck}},\ }\bibfield  {title} {\bibinfo {title} {Exact critical exponents
  for quantum spin chains, non-linear $\sigma$-models at $\theta$= $\pi$ and
  the quantum hall effect},\ }\href
  {https://www.sciencedirect.com/science/article/abs/pii/0550321386901677}
  {\bibfield  {journal} {\bibinfo  {journal} {Nucl. Phys. B}\ }\textbf
  {\bibinfo {volume} {265}},\ \bibinfo {pages} {409} (\bibinfo {year}
  {1986})}\BibitemShut {NoStop}%
\end{thebibliography}%
%------------------------------------------------------------------------------------------------------------------------------------------------------------------------------------------------------------
\end{document}